\documentclass[aps,twocolumn,amsmath,amssymb,reprint,superscriptaddress,notitlepage]{revtex4-1}
\usepackage{graphicx}
\usepackage{latexsym}
\usepackage{times}
\usepackage{color}
\usepackage{amsmath}
\usepackage{dcolumn}
\usepackage{mathbbol}
\usepackage{latexsym,amsmath,amssymb,bm,euscript}
\usepackage{xr}
\usepackage{lipsum}
\usepackage[normalem]{ulem}
\usepackage{hyperref}
\usepackage{esvect}
\usepackage{color} % use colored text in latex

\newcommand{\ket}[1]{| #1 \rangle}
\newcommand{\bra}[1]{\langle #1 |}

\begin{document}
\title{Coherent microwave photon mediated coupling \\between a semiconductor and a superconductor qubit}

\affiliation{Department of Physics, ETH Z\"urich, CH-8093 Z\"urich, Switzerland}
\affiliation{Institut quantique and D\'{e}partment de Physique, Universit\'{e} de Sherbrooke, Sherbrooke, Qu\'{e}bec J1K 2R1, Canada}
\affiliation{Canadian Institute for Advanced Research, Toronto, ON, Canada}
\author{P.~Scarlino$^{1\dagger *}$, D.~J.~van~Woerkom$^{1\dagger}$, U. C. Mendes$^{2}$, J.~V.~Koski$^{1}$, A.~J.~Landig$^{1}$, C.~K.~Andersen$^{1}$, S.~Gasparinetti$^{1}$, C. Reichl$^{1}$, W. Wegscheider$^{1}$, K. Ensslin$^{1}$, T. Ihn$^{1}$, A. Blais$^{2,3}$ and A. Wallraff$^{1}$}

\maketitle
\begin{center}
$^\dagger$ These authors contributed equally to this work.

$^*$ To whom correspondence should be addressed.

\end{center}

\textbf{Semiconductor qubits rely on the control of charge and spin degrees of freedom of electrons or holes confined in quantum dots (QDs). They constitute a promising approach to quantum information processing \cite{Veldhorst2015,Watson2017a}, complementary to superconducting qubits \cite{wendin2017}.
Typically, semiconductor qubit-qubit coupling is short range \cite{Nowack2011,Shulman2012,Veldhorst2015,Watson2017a}, effectively limiting qubit distance to the spatial extent of the wavefunction of the confined particle, which represents a significant constraint towards scaling to reach dense 1D or 2D arrays of QD qubits.
Following the success of circuit quantum electrodynamics \cite{wallraff2004}, the strong coupling regime between the charge \cite{Stockklauser2017,Mi2017} and spin \cite{Mi2017d,Landig2017,Samkharadze2017} degrees of freedom of electrons confined in semiconducting QDs interacting with individual photons stored in a microwave resonator has recently been achieved.
In this letter, we demonstrate coherent
coupling between a superconducting transmon qubit and a semiconductor double quantum dot (DQD) charge qubit mediated by virtual microwave photon excitations in a tunable high-impedance SQUID array resonator acting as a quantum bus
\cite{Blais2007,Majer2007,Sillanpaa2007}.
The transmon-charge qubit coherent coupling rate ($ \sim 21$ MHz)
exceeds the linewidth of both the transmon ($\sim 0.8$ MHz) and the DQD charge ($\sim 3$ MHz) qubit.
By tuning the qubits into resonance for a controlled amount of time, we observe coherent oscillations between the constituents of this hybrid quantum system.
These results
enable a new class of experiments exploring
the use of the two-qubit interactions mediated by microwave photons to create entangled states between semiconductor and superconducting qubits.
The methods and techniques presented here are transferable to QD devices based on other material systems and can be beneficial for spin-based hybrid systems.}

Single electron spins confined in semiconductor quantum dots (QDs) can preserve their coherence for hundreds of microseconds in $^{28}$Si \cite{Muhonen2014,Veldhorst2014} and have typical relaxation times of seconds \cite{Watson2017,Morello2010}. This property can be explored, for example, to build memories for quantum information processors in hybrid architectures combining superconducting qubits and spin qubits.
Strategies to interconnect semiconductor qubits include the control of short-range interactions through the direct overlap of electronic wavefunctions \cite{Nowack2011,Veldhorst2015}, the direct capacitive coupling between QDs \cite{Shulman2012}, enhanced by floating metallic gates \cite{trifunovic2012}, shuttling of electrons between distant QDs by surface acoustic waves \cite{mcneil2011,hermelin2011}, by time-varying gate voltages \cite{Baart2016b} and by fermionic cavities \cite{nicoli2017}.
An alternative approach which allows for long-range qubit-qubit interaction, inspired by superconducting circuit quantum electro-dynamics (QED) and recently explored also for semiconductor QDs \cite{Frey2012,Petersson2012a,Delbecq2011}, is to use microwave photons confined in superconducting resonators to mediate coupling between distant qubits. In this approach, the microwave resonator not only acts as a quantum bus, but also allows for quantum non-demolition qubit readout \cite{Blais2007,Majer2007,Sillanpaa2007}.

With the well established strong coupling of superconducting qubits to microwave resonators \cite{wallraff2004} and the recently achieved strong coupling to charge states in semiconductor double dot structures \cite{Stockklauser2017,Mi2017}, it is now possible to create a microwave photon-based interface between superconducting and semiconducting qubits mediated by a joint coupling resonator.
A similar strategy has been explored in hybrid structures interfacing a transmon qubit with excitations of a spin-ensemble of NV centers in diamonds  \cite{Kubo2011,Kubo2012,Grezes2014} and of collective spins (magnons) in ferromagnets \cite{Tabuchi2015,tabuchi2016,Lachance-Quirion2017}. Furthermore, direct coupling between a superconducting flux qubit and an electron spin ensemble in diamond was investigated \cite{zhu2011c}.
In these works the strong coupling regime was achieved with ensembles, for which the coupling strength scales with the square root of the number of two-level systems interacting with the resonator mode.

\begin{figure*}
\includegraphics[width=1.0\textwidth]{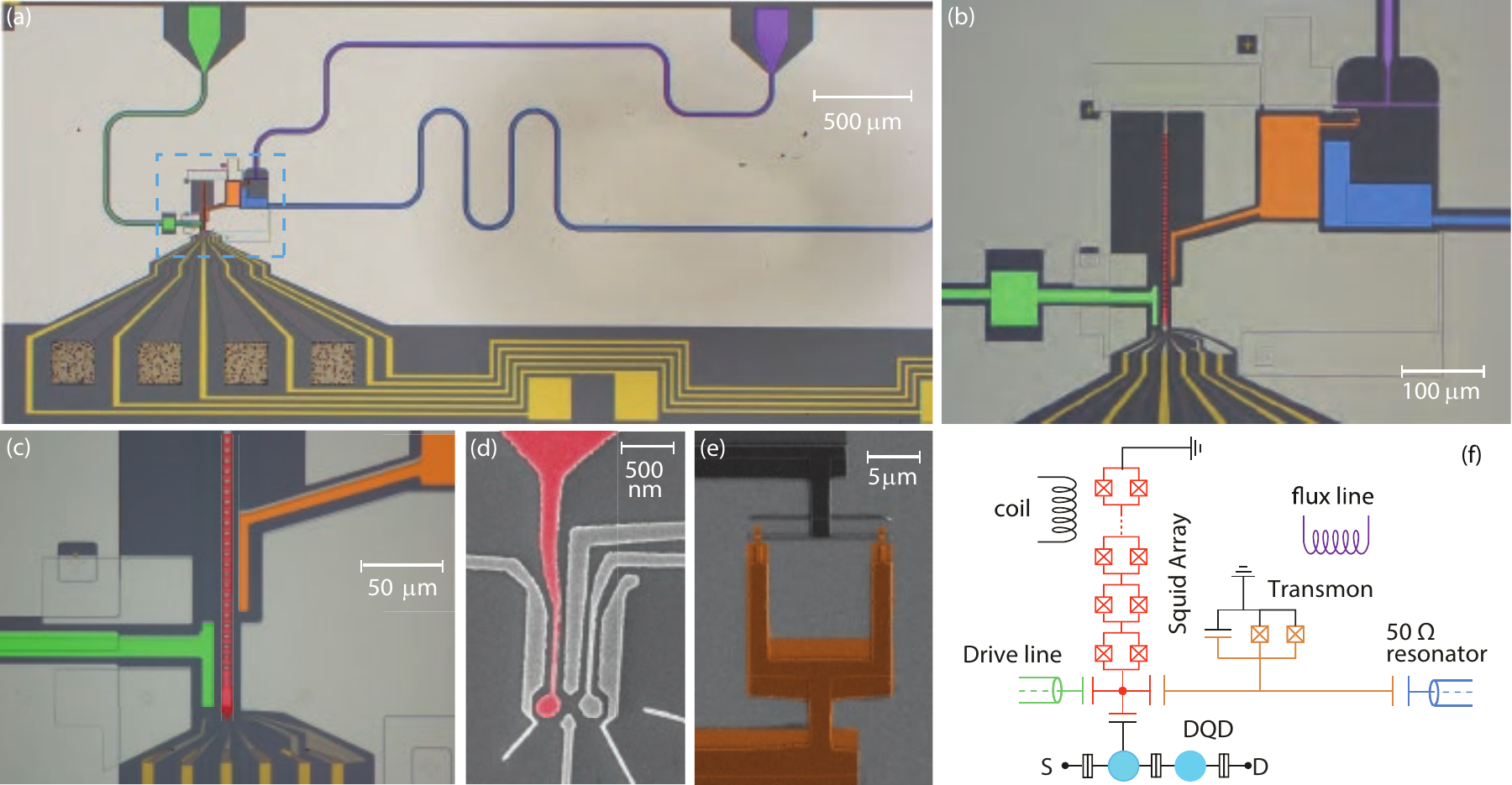} 
\caption{
Sample and simplified circuit diagram. (a) False color optical micrograph of the device showing the substrate (dark gray), the Al superconducting structures forming the groud plane (light gray), the DQD Au gate leads (yellow), the SQUID array resonator (red), its microwave feedline (green), the single island transmon (orange), its readout 50 $\Omega$ coplanar waveguide resonator (blue) and the flux line (purple). (b) Enlarged view of the sample area enclosed by the blue dashed line in panel (a).
(c) Enlarged view of the coupling side of the SQUID array.
(d) Electron micrograph of the DQD showing its electrostatic top gates (Al-light gray) and the plunger gate coupled to the SQUID array (red).
(e) Electron micrograph of the transmon SQUID.
(f) Circuit diagram schematically displaying the DQD [with its source ($S$) and drain ($D$) contact], capacitively coupled to the SQUID array resonator, which in turn is coupled to the transmon. The transmon and the SQUID array are respectively capacitively coupled to a 50 $\Omega$ CPW resonator and microwave feedline. Their resonance frequencies can be tuned by using a flux line and a coil schematically shown in the circuit diagram. The color code is consistent with the optical micrographs.
}
\label{fig:SampleAndCircuit}
\end{figure*}

Here, we explore the coupling of the charge degree of freedom of a single electron confined in a double quantum dot (DQD) to a superconducting transmon qubit in the circuit QED architecture \cite{wallraff2004}.
To perform our experiments, we integrate four different quantum systems into a single device: a semiconductor DQD charge qubit, a superconducting qubit, and two superconducting resonators [see Fig.~\ref{fig:SampleAndCircuit}(a)].
One resonator acts as a quantum bus between the superconducting and the semiconductor qubits and the other one as a readout resonator for the superconducting qubit.
In this way, the functionality for qubit readout and coupling is implemented using two independent resonators at different frequencies, allowing for more flexibility in the choice of coupling parameters and reducing unwanted dephasing due to residual resonator photon population \cite{Schuster2005}.
A simplified circuit diagram of the device is shown in Fig.~\ref{fig:SampleAndCircuit}(f).
The superconducting qubit is of transmon type and consists of a single superconducting aluminum (Al) island shunted to ground via a SQUID (orange in Fig.~\ref{fig:SampleAndCircuit}). The transmon charging and Josephson energies are $E_\mathrm{c}/h \sim 243\, \mathrm{MHz}$ and $E_\mathrm{J}^0/h \sim 30\, \mathrm{GHz}$, respectively (see Methods for more information).
The transition frequency $\omega_\mathrm{tr}$ between its ground state $\ket{g}$ and excited state $\ket{e}$ is adjusted by using the magnetic flux generated in the transmon SQUID loop by a flux line (purple in Fig.~\ref{fig:SampleAndCircuit}).
We read out the state of the transmon qubit with a $50\, \Omega$  coplanar waveguide resonator (dark blue in Fig.~\ref{fig:SampleAndCircuit}) capacitively coupled to the qubit
\cite{wallraff2004,wallraff2005}.

The DQD charge qubit [Fig.~\ref{fig:SampleAndCircuit}(d)], schematically indicated by the 2 light blue dots in Fig.~\ref{fig:SampleAndCircuit}(f), is defined by standard depletion gate technology using Al top gates on a GaAs/AlGaAs heterostructure that hosts a two-dimensional electron gas (2DEG)
\cite{Frey2012,Stockklauser2017,Scarlino2017b}. The DQD is tuned to the few-electron regime and its excitation
energy is given by $ \omega_\text{DQD}=\sqrt{4t_\text{c}^2+\delta^2}$, with the inter-dot tunnel rate $t_\text{c}$ and the DQD energy detuning $\delta$.

We use a superconducting high-impedance resonator for mediating interactions between the transmon and the DQD \cite{Stockklauser2017}.
The resonator is composed of an array of $35$ SQUIDs [Fig.~\ref{fig:SampleAndCircuit}(b)] and is capacitively coupled to both transmon and DQD charge qubits [see Fig.~\ref{fig:SampleAndCircuit}(b,f)].
It is grounded at one end and terminated in a small island at the other end to which a single coplanar drive line is capacitively coupled [green in Fig.~\ref{fig:SampleAndCircuit}(b,c)]. A gate line extends from the island and forms one of the plunger gates of the DQD [in red in Fig.~\ref{fig:SampleAndCircuit}(d)] \cite{Stockklauser2017,Scarlino2017b}.
The high impedance of the resonator increases the strength of the vacuum fluctuations of electric field, enhancing the coupling strength of the individual qubits to the resonator (see Methods for more information).

We characterize the hybrid circuit by measuring the amplitude and phase change of the reflection coefficient of a coherent tone at frequency $\omega_p$ reflected from the multiplexed
resonators (the microwave setup is presented in Extended Data Fig.~\ref{fig:Simplified_circuit}). The response changes with the potentials applied to the gate electrodes forming the DQD and the magnetic flux applied to the transmon.
By varying the DQD detuning $\delta$ and the transmon flux $\Phi_\text{tr}$, each qubit is individually tuned into resonance with the high-impedance resonator.
The coupling strengths measured between the SQUID array resonator and the DQD charge qubit and the transmon qubit are $2g_\text{DQD,Sq}/2\pi \sim 66$ MHz (at $\omega_\text{r,Sq}/2\pi = 4.089$ GHz) and $2g_\text{tr,Sq}/2\pi \sim 451$ MHz (at $\omega_\text{r,Sq}/2\pi = 5.18$ GHz), respectively, for more details see Methods and Extended Data Fig.~\ref{fig:SampleAndCircuit3b}.
For the same configuration, we extract the linewidth of the qubits
spectroscopically \cite{Schuster2005,Stockklauser2017} and find $\delta \omega_{\rm{DQD}}/ 2 \pi \sim 3 \,\rm{MHz}$ and $\delta \omega_{\rm{tr}}/ 2 \pi \sim0.8 \,\rm{MHz}$.
Both subsystems individually are in the strong coupling regime ($2g > \kappa/2 + \gamma_2$) with a SQUID array resonator linewidth of $\kappa / 2 \pi= (\kappa_\text{ext} + \kappa_\text{int})/ 2 \pi \sim (3+5) \,\rm{MHz}$.
\begin{figure}[ht]
\begin{center}

\includegraphics[width=0.50\textwidth]{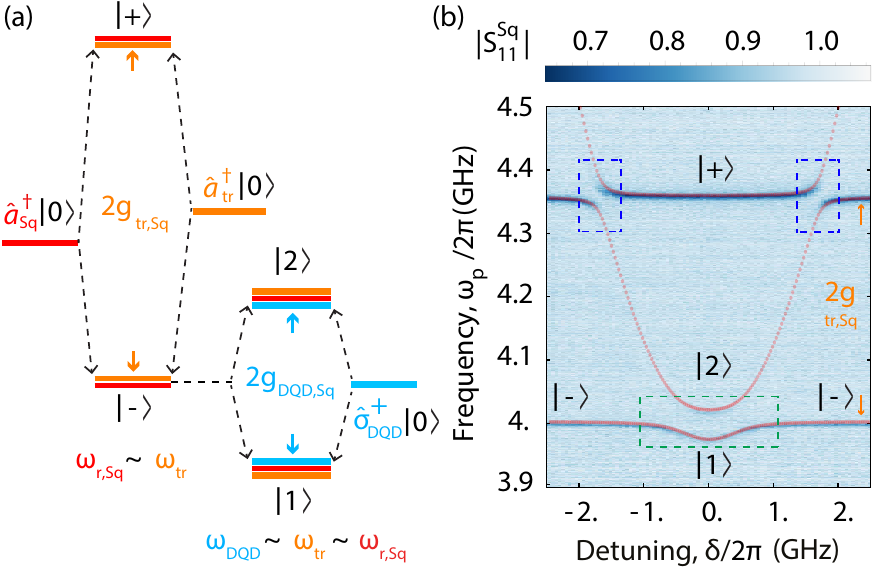}
\end{center}
\caption{
Resonant interaction between the DQD charge qubit, the SQUID array resonator and the transmon.
(a) Energy level diagram of the DQD-SQUID array-transmon system for the bias point considered in panel (b).
The energy levels are colored in accordance with the code used in Fig.~\ref{fig:SampleAndCircuit}.
(b) Reflectance $|S_{11}|$ of the SQUID array resonator hybridized with the transmon and DQD as a function of the DQD detuning $\delta$ at the bias point discussed in the main text.
Red dots are
obtained by numerical diagonalization of the system Hamiltonian [see Eq.~\eqref{hamilt} in Methods],  using parameters extracted from independent spectroscopy measurements.
}
\label{fig:Schematic}
\end{figure}

To demonstrate the coherent coupling between the transmon qubit and the DQD charge qubit, we first characterize the configuration with the three systems interacting  resonantly with each other (see Fig.~\ref{fig:Schematic}).
We tune the SQUID array into resonance with the transmon and observe the vacuum Rabi modes
$\ket{\mp}= \left( \sin \theta_\text{m} \hat{a}^\dagger_\text{Sq} \pm \cos \theta_\text{m} \hat{a}^\dagger_\text{tr} \right)  \ket{0}$,
with the ground state of the system $\ket{0}=\ket{0}_\text{Sq} \otimes \ket{g}_\text{tr} \otimes \ket{g}_\text{DQD}$
and the creation operators for the excitations in the SQUID array (transmon) $\hat{a}^\dagger_\text{Sq}$ ($\hat{a}^\dagger_\text{tr}$).
The mixing angle $\theta_\text{m}$ is determined by $\tan 2\theta_\text{m} = 2g_\text{tr,Sq}/ |\Delta_\textrm{tr}|$, with $|\Delta_\textrm{tr}|= |\omega^\prime _{\rm{tr}}-\omega_{\rm{r,Sq}}|$
and the transmon excitation frequency $\omega_\text{tr}^\prime$ dressed by the interaction with the 50 $\Omega$ resonator.
We then configure the DQD electrostatic gate voltages to tune its transition frequency
at the charge sweet spot [$\omega_\text{DQD}(\delta = 0)=2t_\text{c}$] into resonance with the lower transmon-SQUID array Rabi mode $\ket{-}$.
From the hybridization between the states $\ket{-}$ and the DQD excited state $ \hat{\sigma}^+_\text{DQD}\ket{0}$, we obtain the states $\ket{1}$ and $\ket{2}$,
leading to the avoided crossing indicated by the green dashed box in Fig.~\ref{fig:Schematic}(b).
Similarly, when the DQD excitation energy is equal to the energy of the higher transmon-SQUID array Rabi mode, $\ket{+}$, the hybrid system develops two avoided crossings at the respective detunings $\delta$ in the spectrum [see blue dashed box in Fig.~\ref{fig:Schematic}(b)].
The observed spectrum resulting from the hybridization of the three quantum systems is in good agreement with our calculation (see 'System Hamiltonian' section in Methods) [red dots in  Fig.~\ref{fig:Schematic}(b)].

\begin{figure*}[ht]
\begin{center}
\includegraphics[width=\textwidth]{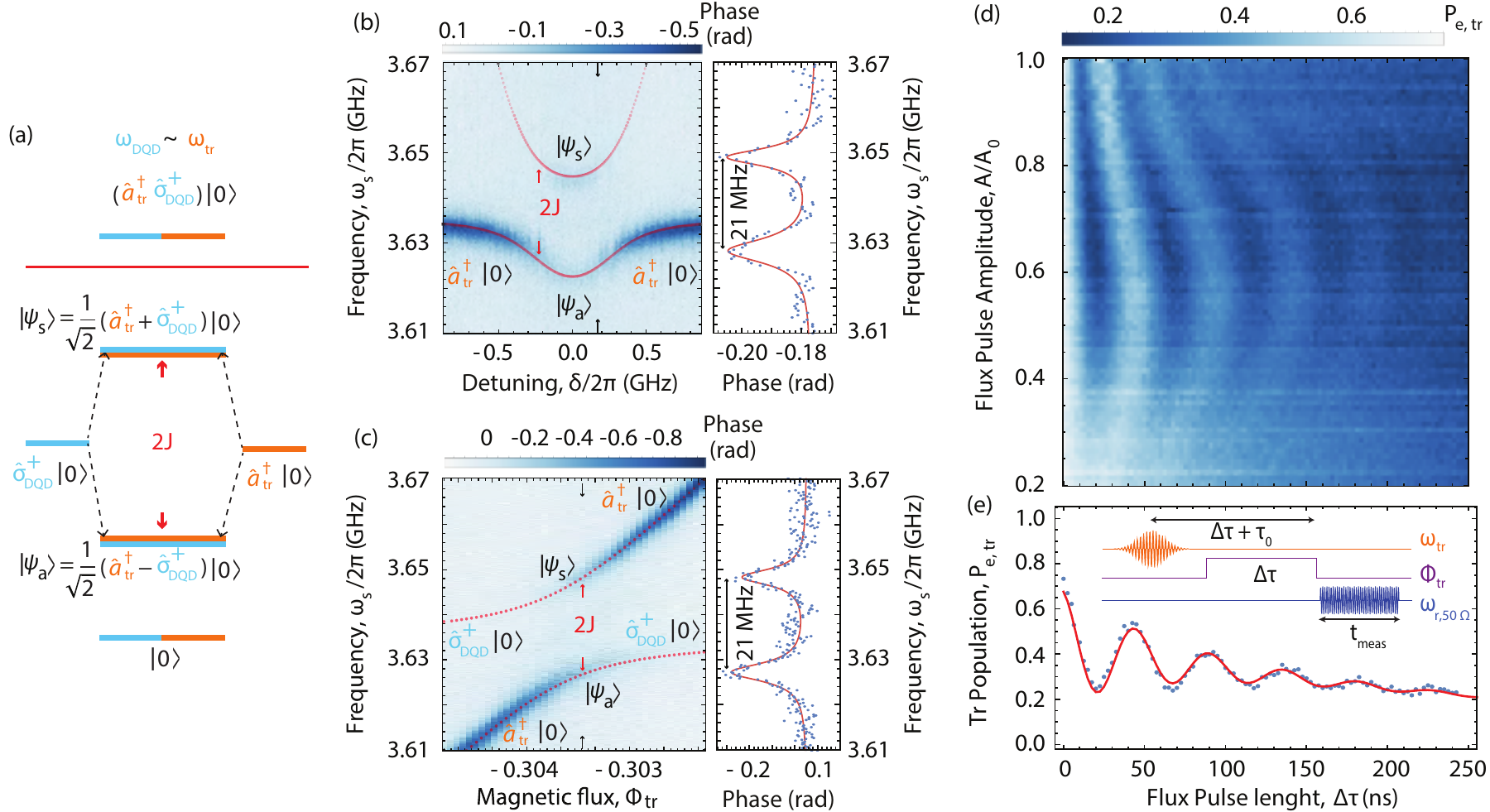} 
\end{center}
\caption{
DQD-transmon interaction mediated by virtual photon exchange in the SQUID array resonator.
(a) Energy level diagram of the DQD-transmon qubit coupling mediated via dispersive interaction with the SQUID array resonator (red line). The DQD excitation ($\sigma^\dagger_{\text{DQD}} \ket{0}$) and the transmon excitation ($a^\dagger_{\text{tr}} \ket{0}$) are shown, together with their hybridized states $\ket{\Psi_{\text{s,a}}}$, the system vacuum state $\ket{0}=\ket{0}_\text{Sq} \otimes \ket{g}_\text{tr} \otimes \ket{g}_\text{DQD}$ and the doubly excited states $a^\dagger_{\text{tr}} \sigma^\dagger_{\text{DQD}} \ket{0}$.
(b) Left: spectroscopy of the DQD qubit interacting with the transmon. Phase $\Delta \phi=\mathrm{Arg}[S_{11}]$ of a fixed frequency measurement tone $\omega_{\rm{p}}/2 \pi =6.5\,\rm{GHz}=\omega_{\rm{r,50\Omega}}/ 2 \pi$ reflected off the 50 $\Omega$ CPW read-out resonator $\it{vs.}$ transmon qubit spectroscopy frequency $\omega_{\rm{s}}$ and DQD qubit detuning $\delta$ [(c) the flux through the SQUID loop of the transmon $\Phi_\text{tr}$].
Right: phase $\Delta \phi=\mathrm{Arg}[S_{11}]$ response at the DQD detuning $\delta$ [(c) at the flux $\Phi_\text{tr}$] indicated by the red arrows in left panel showing a coupling splitting of $2J \sim 21\,\rm{MHz}$.
(d) Population transfer between the transmon and the DQD charge qubit induced by the pulse protocol depicted in the inset of panel (e).
Average transmon excited state population $P_{\rm{e,tr}}$ (each data point is the intergrated average over 50000 repetitions of the experiment), as a function of the flux pulse length $\Delta \tau$ and normalized flux pulse amplitude $A/A_0$.
(e) Transmon excited state population $P_{\rm{e,tr}}$ vs.~$\Delta \tau$ for a flux pulse amplitude of $A/A_0=0.6$, for which the transmon is approximately in resonance with the DQD ($\omega_{\rm{tr}}/2 \pi \sim \omega_{\rm{DQD}}/2 \pi=3.660\,\rm{GHz}$). $\omega_{\rm{r,Sq}} /2 \pi = 4.060\,\rm{GHz}$. The red line is a fit to a Markovian master equation model (see Methods for details).
}
\label{fig:SampleAndCircuit5}
\end{figure*}

Next, we discuss the virtual photon-mediated coherent interaction between the DQD and the transmon qubit.
This is realized in the dispersive regime, where both qubit frequencies are detuned from the high-impedance resonator.
In this regime, no energy is exchanged between the qubits and the resonator, and the strength of the effective coherent interaction between the two qubits is given by
$2J \sim g_{\rm{tr,Sq}}g_{\rm{DQD,Sq}}/(1/|\Delta_\textrm{tr}| + 1/|\Delta_\textrm{DQD}| )$ \cite{Majer2007,Sillanpaa2007}.
We spectroscopically explore this qubit-qubit coupling by applying a probe tone at frequency $\omega_\text{p}/2\pi = 6.55\, \rm{GHz}$ to the transmon readout resonator. The reflectance of the probe tone from the $50\, \Omega$ resonator
is measured while a microwave spectroscopy tone of frequency $\omega_\text{s}$ is swept across the transmon transition frequency to probe its excitation spectrum \cite{Schuster2005}.

To observe the coherent DQD-transmon coupling, we either tune $\delta$ to bring the DQD into resonance with the transmon [see Fig.~\ref{fig:SampleAndCircuit5}(b)] or tune $\Phi_\text{tr}$ to bring the transmon into resonance with the DQD [see Fig.~\ref{fig:SampleAndCircuit5}(c)].
In either case, when the qubit frequencies are in resonance, as depicted in Fig.~\ref{fig:SampleAndCircuit5}(a), a clear avoided crossing of magnitude $2J/2\pi \sim 21\, \rm{MHz}$, larger than the combined linewidth of the coupled system $(\gamma_{\rm{DQD}}  + \gamma_{\rm{tr}})/2\pi \sim 4 \,\rm{MHz}$, is observed [Figs.~\ref{fig:SampleAndCircuit5}(b,c)].
The observed resonance frequencies are in good agreement with our simulation [red dots in Fig.~\ref{fig:SampleAndCircuit5}(b,c)] for the explored configuration characterized by $|\Delta_{\rm{DQD}}| \sim \, 10 g_{\rm{DQD,Sq}}$, $|\Delta_\textrm{tr}|  \sim  \,3 g_{\rm{tr,Sq}}$.
The well-resolved DQD-transmon avoided crossing demonstrates that the high-impedance resonator mediates the coupling between the semiconductor and the superconductor qubit.

We demonstrate virtual-photon mediated coherent population transfer between the transmon and DQD charge qubits in time-resolved measurements. We induce the exchange coupling by keeping
the DQD and SQUID array cavity frequencies fixed at $\omega_{\rm{DQD}} (\delta=0)/2\pi=3.66 \, \rm{GHz}$ and $\omega_{\rm{r,Sq}}/2\pi=4.06 \, \rm{GHz}$, respectively, and varying the transmon frequency non adiabatically
\cite{Blais2007,Majer2007}
using the pulse protocol illustrated in the inset of Fig.~\ref{fig:SampleAndCircuit5}(e).
Initially, both qubits are in their ground state and the effective coupling between them is negligible, due to the large difference between their excitation frequencies.
Next, we apply a $\pi$-pulse to the transmon qubit to prepare it in its excited state.
Then, a non-adiabatic current pulse, applied to the flux line, changes the flux $\Phi_\text{tr}$ and tunes the transmon into resonance with the DQD charge qubit for a time $\Delta \tau$, which we vary between 0 and 250 ns.
After the completion of the flux pulse controlling the interaction, the state of the transmon is measured through its dispersive interaction with the 50 $\Omega$ CPW resonator.
We observe coherent oscillations of the transmon excited state population as a function of the interaction time $\Delta \tau$ [see Figs.~\ref{fig:SampleAndCircuit5}(d,e)].
The frequency of these oscillations, Fig.~\ref{fig:SampleAndCircuit5}(d), is determined by the interaction strength $J(|\Delta_\textrm{tr}|,|\Delta_\textrm{DQD}|)$.
As a result, we observe the characteristic chevron pattern in the transmon qubit population in dependence on the flux pulse amplitude and length [see Fig.~\ref{fig:SampleAndCircuit5}(d)] \cite{Blais2007}.

A trace of the population oscillation pattern at fixed pulse amplitude $A/A_0 \sim 0.6$, approximately realizing the DQD-transmon resonance condition ($\omega^\prime_{\rm{tr}}/ 2 \pi =\omega_{\rm{DQD}}/ 2 \pi=3.66\,\rm{GHz}$), is in excellent agreement with the Markovian master equation simulation [see the red line in Fig.~\ref{fig:SampleAndCircuit5}(e) and Methods for more information].
The simulations are performed within the dispersive approximation in which the qubits interact with rate $2J/2\pi = 21.6\,\text{MHz}$, via an exchange interaction
consistent with the spectroscopically measured energy splitting $2J/2\pi \sim 21\, \rm{MHz}$ [Fig.~\ref{fig:SampleAndCircuit5}(b-e)].

In this work, we realized an interface between semiconductor- and superconductor-based qubits by exchanging virtual photons between two distinct physical systems in a hybrid circuit QED architecture \cite{Xiang2013a,Kurizkia2015}. The coherent interaction between the qubits is witnessed both by measurements of well-resolved spectroscopic level splitting and by time-resolved population oscillations. The interaction can be enabled both electrically via the quantum dot and magnetically via the transmon qubit. The resonator mediated coupling also provides for non-local coupling to the semiconductor qubit, demonstrated here over distances of more than $50 \,\rm{\mu m}$. We expect the approach demonstrated here for the charge degree of freedom of the semiconductor qubit to be transferable to the spin degree of freedom and also to other material systems such as Si or SiGe \cite{Mi2017d,Landig2017,Samkharadze2017}. In this way, the coupling to electron spin or even nuclear spin qubits may provide an avenue for realizing a spin based quantum memory, which can be interfaced to other solid state qubits, including superconducting ones. In addition, the combination of short distance coupling and control in semiconductor qubits with long-distance coupling through microwave resonators provided by circuit QED may indicate a viable solution to the wiring and coupling challenge in semiconductor qubits \cite{Vandersypen2017} and may be essential for realizing error correction in these systems, for example by using the surface code \cite{Fowler2012}.\\

\textbf{Acknowledgment}\\
We would also like to thank Agustin Di Paolo, Michele Collodo, Philipp Kurpiers, Johannes Heinsoo and Simon Storz for helpful discussions and for valuable contributions to the experimental setup, software and numerical simulations.
This work was supported by the Swiss National Science Foundation (SNF) through the National Center of Competence in Research (NCCR) Quantum Science and Technology (QSIT), the project Elements for Quantum Information Processing with Semiconductor/Superconductor Hybrids (EQUIPS) and by ETH Zurich.
UCM and AB were supported by NSERC and the Canada First Research Excellence fund.\\

%Author contributions
\textbf{Author Contributions}\\
PS designed the sample with inputs from SG, DJvW and AW.
PS, DJvW and JVK fabricated the device.
DJvW and PS performed the experiments.
PS and DJvW analysed the data.
UCM, CKA and AB performed the theoretical simulations.
CR and WW provided the heterostructure.
PS, UCM and AW wrote the manuscript with the input of all authors.
AW, KE and TI coordinated the project.\\

\textbf{Data and materials availability:} The data presented in this paper and corresponding supplementary material are available online at ETH Zurich repository for research data, https://www.research-collection.ethz.ch/.
\vspace{0.2 cm}
\bibliographystyle{naturemag}
\bibliography{ReferenceDatabase}

\setlength{\parindent}{0pt}

\appendix

\setlength{\parindent}{2pt}

\section*{Methods}

\setlength{\parindent}{15pt}

\addtocontents{toc}{\setcounter{tocdepth}{0}}

\subsection*{Device and measurement setup}
The device is realized on a GaAs/Al$_\mathrm{x}$Ga$_\mathrm{1-x}$As heterostructure.
The 2DEG, embedded 90 nm below the surface at the interface of GaAs/Al$_\mathrm{x}$Ga$_\mathrm{1-x}$As, has been removed by etching everywhere but in a small region hosting the DQD [see Figs.~\ref{fig:SampleAndCircuit}(a,b)].
The gate structures that define the DQD confinement potential are realized using a combination of gold (Au) top gates for the coarse gate structures [yellow in Figs.~\ref{fig:SampleAndCircuit}(a-c)], and  aluminum (Al) for the fine gate structures [light gray
in Figs.~\ref{fig:SampleAndCircuit}(a,b)].
The tunnel junctions of the SQUIDs are formed by two Al electrodes separated by a thin oxide layer. They are fabricated using standard electron-beam lithography and shadow evaporation of 30 nm and 120 nm aluminum (with in-situ oxidation).

We characterize the hybrid circuit by measuring the amplitude and phase change of the reflection coefficient of a coherent tone reflected at frequency $\omega_\text{p}$ off the multiplexed SQUID array and $50\, \Omega$ CPW resonators (see Extended Data Fig.~\ref{fig:Simplified_circuit}). The multiplexing of the two resonators is realized by cascading two circulators and connecting the reflection port of each resonator to a circulator (see Extended Data Fig.~\ref{fig:Simplified_circuit}).
The microwave tone is generated at room-temperature and is attenuated by -20~dB at the 4~K, 100~mK and 20~mK stages before passing through a circulator which routes it to the resonator and routes the reflected signal to the output line. In the output line the reflected signal is amplified using a cryogenic HEMT (+39~dB) at 4~K and by two amplifiers (+33~dB each) at room-temperature, before it is down converted to an intermediated frequency (IF) of 250~MHz. With +29~dB amplification the IF is acquired at 1Gs/s using an Acqiris U1084A PCIe 8-bit High-Speed Digitizer.
The DC voltages are supplied to the gates by Yokogawa 7651 DC programmable sources with a 1:10 voltage divider also acting as a low pass filter (1~Hz cut-off). The source and drain of the DQD were grounded in the experiment. At base temperature, 2-stage RC filters with 160~kHz and 16~kHz cut-off are used at the input of shielded lines leading to the sample holder.
A schematic of the complete setup with all important components is displayed in Extended Data Fig.~\ref{fig:Simplified_circuit}. The experiment is performed at a cryostat temperature of 30 mK.

\setcounter{figure}{0}
\renewcommand{\figurename}{Extended Data Figure}

\begin{figure}[!h]
\begin{center}
\includegraphics[width=0.5\textwidth]{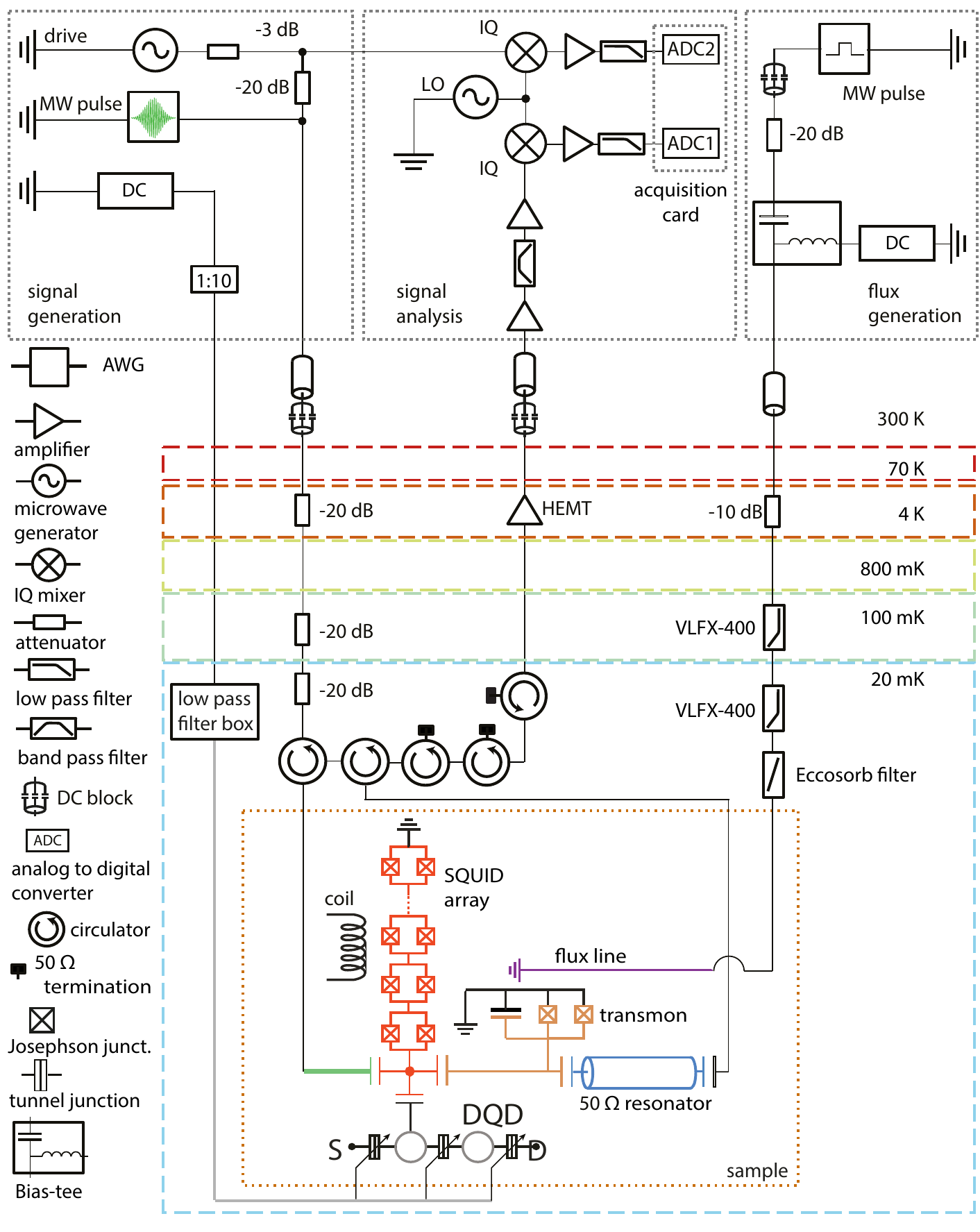}
\end{center}
\caption{Simplified schematic of cryogenic and room-temperature measurement setup, further details in text. }
\label{fig:Simplified_circuit}
\end{figure}

\subsection*{System Hamiltonian}

The coupled quantum system is described by the Hamiltonian
\begin{align} \label{hamilt}
\hat{H}_\text{tot}=&\hat{H}_\text{DQD} +\hat{H}_\text{tr} +\hat{H}_\text{r, Sq} +\hat{H}_\text{r, 50 $\Omega$} + \\
&+ \hat{H}_\text{DQD, Sq} + \hat{H}_\text{tr, Sq} + \hat{H}_\text{tr, 50 $\Omega$}, \nonumber
\end{align}
\begin{align} \label{hamilt2}
&\hat{H}_\text{DQD} =\frac{\omega_\text{DQD}{(t_\text{c}, \delta)}}{2} \sigma_\text{DQD}^z, \, \hat{H}_\text{tr} =  \sum_{i=1}^{n_\text{tr}}\omega_{i,\text{tr}}(\Phi_\text{Sq},\Phi_\text{tr}) |i\rangle \langle i | ,\\
%&\hat{H}_\text{tr} =  \sum_{i=1}^{n_\text{tr}}\omega_{i,\text{tr}}(\Phi_\text{Sq},\Phi_\text{tr})  |i\rangle \langle i | ,\\
&\hat{H}_\text{r, Sq}= \omega_\text{r,Sq}(\Phi_\text{Sq}) \hat{a}^\dagger_\text{Sq} \hat{a}_\text{Sq} , \, \hat{H}_\text{r, 50 $\Omega$}=\omega_\text{r,50$\Omega$} \hat{b}^\dagger \hat{b} ,\\
%&\hat{H}_\text{r, 50 $\Omega$}=\omega_\text{r,50$\Omega$} \hat{b}^\dagger \hat{b} ,\\
&\hat{H}_\text{DQD, Sq} =g_\text{DQD,Sq}(\Phi_\text{Sq}, t_\text{c}, \delta)(\sigma^-_\text{DQD} \hat{a}_\text{Sq}^\dagger  + \sigma^+_\text{DQD}\hat{a}_\text{Sq} ),\\
&\hat{H}_\text{tr, Sq}=\sum_{i,j=1}^{n_\text{tr}} g_\text{tr,Sq}(\Phi_\text{Sq},\Phi_\text{tr})  n_{i,j} |i\rangle \langle j |(\hat{a}^\dagger_\text{Sq}  + \hat{a}_\text{Sq} ) ,\\
&\hat{H}_\text{tr, 50$\Omega$}=\sum_{i,j=1}^{n_\text{tr}} g_\text{tr,50$\Omega$} (\Phi_\text{Sq},\Phi_\text{tr}) n_{i,j} |i\rangle \langle j | (\hat{b}^\dagger  + \hat{b} ) ,
\end{align}
with $\hbar =1$, and $\hat{a}_\text{Sq}$, $\hat{b}$ and $\hat{\sigma}^-_\text{DQD}$ are the annihilation operator for the excitations of the SQUID array, $50\,\Omega$ resonator and lowering operator of the DQD qubit, respectively. $\omega_\text{r,Sq}$ and $\omega_{\rm{r},50\Omega}$ are the resonance frequency of the SQUID array and of $50\,\Omega$ CPW resonator, respectively.
The DQD charge qubit energy is given by $ \omega_\text{DQD}=\sqrt{4t_\text{c}^2+\delta^2}$, where $t_\text{c}$ and $\delta$ are the inter-dot tunnel rate and DQD energy detuning, respectively.
$\omega_{i,\text{tr}}(\Phi_\text{tr})$ and $|i\rangle$ are the frequency and state of the $i$-level of the transmon, respectively.  $n_{i,j} = \langle i | \hat{n} | j \rangle $ are the Cooper pair number matrix elements, and $n_\text{tr}$ is the number of levels forming the transmon qubit (in our model $n_\text{tr}=4$) \cite{Koch2007}.
The coupling strengths between the transmon-SQUID array, transmon-50$\Omega$ resonator, and DQD-SQUID array are indicated with $g_\text{tr,Sq}(\Phi_\text{Sq},\Phi_\text{tr})$, $g_\text{tr,50$\Omega$}(\Phi_\text{tr})$ and $g_\text{DQD,Sq}(\Phi_\text{Sq}, t_\text{c}, \delta)$, where $\Phi_\text{Sq}$ and $\Phi_\text{tr}$ are the external magnetic fluxes through the SQUID loops of the resonator array (assumed uniformly threaded) and transmon, respectively.

\subsection*{SQUID array resonator}

The SQUID array resonator is formed by $N_\text{Sq}=35$ SQUIDs in series with extra $N_\text{sj}=34$ single junctions generated during the shadow evaporation process. The total inductance is
\begin{equation}
L(\Phi_\text{Sq}) = N_\text{Sq}L_\text{Sq}\left( \beta + \frac{1}{|\cos(\Phi_\text{Sq}^\prime/\Phi_0)|} \right),
\end{equation}
with the magnetic flux quantum $\Phi_0$, and the total magnetic flux through the SQUID array $\Phi_\text{sq}^\prime=2 \pi \gamma \Phi_\text{Sq} + 2 \pi \Phi_c$. Here, $\gamma$ and $\Phi_c$ are constants that were parameterized by fitting Extended Data Fig.~\ref{fig:SampleAndCircuit2}(b).
$\beta = N_\text{sj}L_\text{sj}/N_\text{Sq}L_\text{Sq}$ is a parameter that takes into account the presence of an extra constant inductive contribution, in series with the SQUID array inductance, generated during the shadow mask deposition process \cite{Stockklauser2017,Scarlino2017b}. In our experiment $\beta \sim 0.1$.
In a lumped element model, the SQUID array frequency is
\begin{equation}
\omega_\text{r,Sq}(\Phi_\text{Sq}) = \frac{\omega_\text{r,Sq}^0}{\left( \beta + \frac{1}{|\cos(\Phi_\text{Sq}^\prime/\Phi_0)|} \right)^{1/2}},\label{eq8}
\end{equation}
with $\omega_\text{r,Sq}^0 = 1/\sqrt{C N_\text{Sq} L_\text{Sq}}$ and $C$ an effective capacitance that takes into account the capacitance of the SQUID array to the ground and to the transmon.

\subsection*{Transmon qubit}

The transmon qubit in our device is formed by a single island capacitor shunted to ground by two Josephson junctions in a SQUID geometry. We control the qubit frequency by an external magnetic flux $\Phi_\text{tr}$. For symmetric junctions, the Josephson energy is $E_\text{J}(\Phi_\text{Sq},\Phi_\text{tr})  = E_\textrm{J}^0 |\cos(\Phi_\text{tr}^\prime/\Phi_0)|$, with the Josephson energy at zero flux $E_J^0$,
the total magnetic flux through the transmon $\Phi_\text{tr}^\prime= 2\pi \alpha \Phi_\text{Sq} + 2\pi \Phi_\text{tr}$. It depends on both external fluxes $\Phi_\text{tr}$ and $\Phi_\text{Sq}$. Here, $\alpha$ is the ratio between the areas of the SQUID loops of the transmon and the SQUID array resonator. The transmon Hamiltonian can be written as
\begin{equation}
\hat{H}_\text{tr} = 4 E_c (\hat{n} - n_g)^2 -  E_\text{J}(\Phi_\text{Sq},\Phi_\text{tr})\cos \hat{\varphi},
\end{equation}
with the Cooper pair number operator $\hat{n}$, the effective offset  charge $n_g$ of the device, the phase difference $\hat {\varphi}$ across the junction, and the transmon charging energy $E_c$. We approximate the transmon frequency (0 to 1 transition) as \cite{Koch2007}
\begin{equation}
\omega_\text{tr}(\Phi_\text{Sq},\Phi_\text{tr}) = \omega_\text{pl,tr} |\cos(\Phi_\text{tr}^\prime/\Phi_0)|^{1/2} - E_c \label{eq10}
\end{equation}
with the plasma frequency $\omega_\text{pl,tr} = \sqrt{8 E_c E_J^0}$ obtained by fitting the model to the data in Extended Data Fig.~\ref{fig:SampleAndCircuit2}. Finally, the transmon Hamiltonian is diagonalized numerically considering four states.

\subsection*{Double quantum dot qubit}

The Hamiltonian describing the DQD charge qubit is
\begin{equation} \label{eq1}
\hat{H}_\text{DQD} = \frac{\delta}{2}\hat{\tau}^z + t_c \hat{\tau}_x
\end{equation}
where $\delta$ is the detuning between the two dots, $t_c$ is the interdot tunneling coupling, and $\hat{\tau}_i$ are the three Pauli matrices defined in the $|L\rangle$ and  $|R\rangle$ basis, the bases for a single charge to be either on the left or the right QD [$\hat{\tau}_z = |R\rangle \langle R | - |L\rangle \langle L|$].

A basis rotation is performed to diagonalize the DQD Hamiltonian, resulting in
\begin{equation} \label{eq2}
\hat{H}_\text{DQD} = \frac{\omega_\text{DQD}}{2}\sigma_\text{DQD}^z,
\end{equation}
where $\omega_\text{DQD} = \sqrt{\delta^2 + 4 t_c^2}$ and $\sigma_\text{DQD}^z =  \ket{+}_\text{DQD}  \bra{+}_\text{DQD} -  \ket{-}_\text{DQD} \bra{-}_\text{DQD}$, with
\begin{align} \label{eq3}
 \ket{+}_\text{DQD} &= \cos (\theta/2) \ket{R} - \sin (\theta/2) \ket{L} \\
 \ket{-}_\text{DQD} &= \sin (\theta/2) \ket{R} + \cos (\theta/2) \ket{L}
\end{align}
and $\tan \theta = 2 t_c/\delta$.

\subsection*{Description of the strategy used to get the parameters for the modelling}

The coupling strengths between the transmon and SQUID array, the transmon and 50~$\Omega$ resonator, and the DQD and SQUID array are defined as:
\begin{align}
g_\text{tr,Sq}(\Phi_\text{Sq},\Phi_\text{tr}) &=  g_\text{tr,Sq}^0 \frac{|\cos (\Phi_\text{tr}^\prime/\Phi_0)|^{1/4}}{\left(\beta  + \frac{1}{|\cos(\Phi_\text{Sq}^\prime/\Phi_0)|}\right)^{1/4}} \label{eq15},\\
g_{\text{tr},50~\Omega}  (\Phi_\text{Sq},\Phi_\text{tr}) & =  g_\text{tr,50$\Omega$}^0  |\cos (\Phi_\text{tr}^\prime/\Phi_0)|^{1/4}, \\
g_\text{DQD,Sq}(\Phi_\text{Sq},t_\text{c}, \delta) & = \frac{g_\text{DQD,Sq}^0 }{\left(\beta  + \frac{1}{|\cos(\Phi_\text{Sq}^\prime/\Phi_0)|}\right)^{1/4}}\frac{2t_\text{c}}{\omega_\text{DQD}(t_\text{c}, \delta)}. \label{eq17}
\end{align}
Here, $\Phi_\text{Sq}^{\prime}$ and $\Phi_\text{tr}^{\prime}$ are the external (total)
magnetic flux through the SQUID loops of the resonator array and the transmon. The term $2t_\text{c} / \omega_\text{DQD}$ in Eq.~\eqref{eq17} corresponds to the mixing angle renormalization of the DQD-resonator interaction strength \cite{Frey2012}.

\begin{figure}[!b]
\begin{center}
\includegraphics[width=0.50\textwidth]{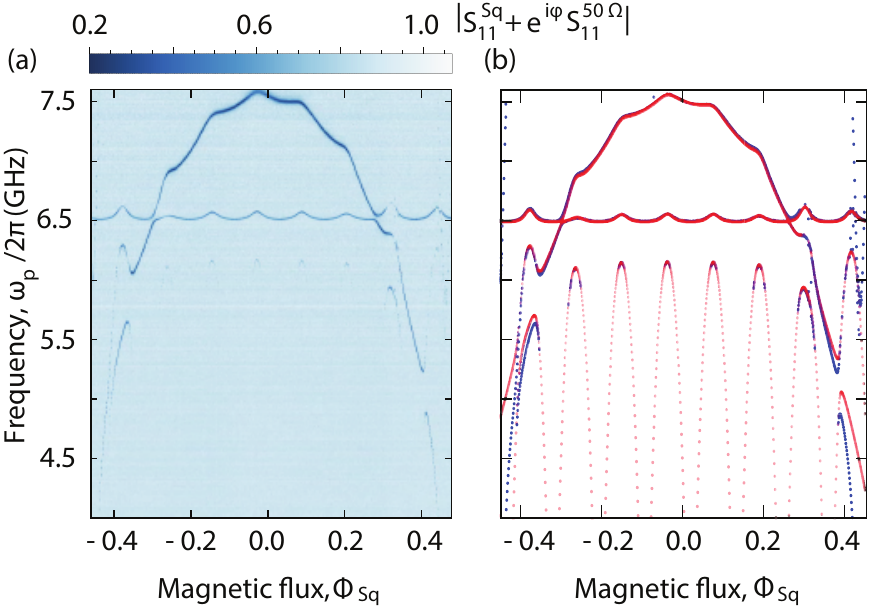}
\end{center}
\caption{
Flux tunability of the SQUID array and Transmon qubit.
(a) Reflectance spectrum $|S_{11}^{\rm{Sq}}(\omega_\text{p})+ e^{i \varphi} S_{11}^{\rm{50 \Omega}}(\omega_\text{p})|$ (with $\varphi$ the phase accumulated by the microwave signal in between the two resonators) of the multiplexed SQUID array and 50 $\Omega$ CPW resonators as a function of probe frequency $\omega_\text{p} / 2\pi$ and applied magnetic flux $\Phi_\text{Sq}$ (expressed in flux quanta for the periodicity of the SQUID array) via the coil.
(b) Resonance frequencies extracted from the dips in the $|S_{11}^{\rm{Sq}}(\omega_\text{p})+ e^{i \varphi} S_{11}^{\rm{50 \Omega}}(\omega_\text{p})|$ (blue points) and simulated spectrum (red points) of the transmon interacting with the SQUID array and the 50 $\Omega$ CPW resonators, according to Eq.~\eqref{hamilt} with parameters
$\omega_{\rm{pl,tr}}/2\pi  = 6.550$ GHz, $E_{\rm{c}}/h = 243$ MHz,
$\omega_{\rm{r,Sq}}^0/2\pi  = 7.867$ GHz, $\omega_{\rm{r},50\Omega}/2\pi  = 6.490$ GHz, $\beta = 0.1$, $g_{\rm{tr,Sq}}^0/2\pi  = 230$ MHz, $g_{\rm{tr},50\Omega}^0/2\pi  = 120$ MHz.
}
\label{fig:SampleAndCircuit2}
\end{figure}

The parameters $g_\text{tr,Sq}^0$ and $g_\text{tr,50$\Omega$}^0 $ are obtained by fitting the experimental data in Extended Data Fig.~\ref{fig:SampleAndCircuit2} (more details are presented in the following section). The DQD-SQUID array coupling in Eq.~\eqref{eq17} is obtained by considering $g_\text{DQD,Sq}(\Phi_\text{Sq}) \propto Z_\text{Sq}^{-1/2}(\Phi_\text{Sq})$, where $Z_\text{Sq}(\Phi_\text{Sq})=\sqrt{L(\Phi_\text{Sq})/C}$ is the SQUID array impedance.
$g_\text{DQD,Sq}^0$ is obtained by fitting the vacuum Rabi splitting data presented in Extended Data Fig.~\ref{fig:SampleAndCircuit3b} with the other parameters fixed.

We calculate the spectrum of the DQD-SQUID array-transmon system by numerical diagonalization of the Hamiltonian \eqref{hamilt} using parameters extracted from independent spectroscopy measurements. The parameters used to obtain theory points in Fig.~\ref{fig:Schematic} and Fig.~\ref{fig:SampleAndCircuit5} are listed in the Tables~\ref{tab:para_fig2}, ~\ref{tab:para_fig3b} and ~\ref{tab:para_fig3c}.

\begin{table}[h!]
\begin{tabular}{|c|c|}
  \hline
  $2t_\text{c}$ & $3993$~MHz  \\ \hline
    $\omega_{\rm{tr}}/ 2 \pi$ & $4150$~MHz \\ \hline
$\omega_{\rm{r,Sq}}/2\pi $ & $4230$~MHz \\ \hline
  $g_{\rm{tr,Sq}}/2\pi$ & $166$~MHz \\ \hline
 $g_\text{DQD,Sq}/2\pi$ & $34$~MHz  \\ \hline
  $g_{\rm{tr},50\Omega}/2\pi $ & $98$~MHz \\
  \hline
\end{tabular}
\caption{Parameters used to obtain the calculated eigenenergies shown in  Fig.~\ref{fig:Schematic}(b).}
\label{tab:para_fig2}
\end{table}

\begin{table}[h!]
\begin{tabular}{|c|c|}
  \hline
  $2t_\text{c}$ & $3635$~MHz  \\ \hline
  $\omega_{\rm{tr}}/ 2 \pi$ & $3695$~MHz \\ \hline
  $\omega_{\rm{r,Sq}}/2\pi $ & $4062$~MHz \\ \hline
  $g_{\rm{tr,Sq}}/2\pi$ & $128$~MHz \\ \hline
  $g_\text{DQD,Sq}/2\pi$ & $36$~MHz  \\ \hline
  $g_{\rm{tr},50\Omega}/2\pi $ & $93$~MHz \\
  \hline
\end{tabular}
\caption{Parameters used to obtain the calculated eigenenergies shown in  Fig.~\ref{fig:SampleAndCircuit5}(b).}
\label{tab:para_fig3b}
\end{table}

\begin{table}[h!]
\begin{tabular}{|c|c|}
  \hline
  $2t_\text{c}$ & $3638$~MHz  \\ \hline
  $\omega_{\rm{tr}}/ 2 \pi$ & $3695$~MHz \\ \hline
  $\omega_{\rm{r,Sq}}/2\pi $ & $4062$~MHz \\ \hline
  $g_{\rm{tr,Sq}}/2\pi$ & $128$~MHz \\ \hline
  $g_\text{DQD,Sq}/2\pi$ & $36$~MHz  \\ \hline
  $g_{\rm{tr},50\Omega}/2\pi $ & $93$~MHz \\
  \hline
\end{tabular}
\caption{Parameters used to obtain the calculated eigenenergies shown in  Fig.~\ref{fig:SampleAndCircuit5}(c).}
\label{tab:para_fig3c}
\end{table}

\subsection*{Interaction between the transmon qubit and the SQUID array and 50~$\Omega$ resonators}

We analyse the flux dependence of the system transition frequencies to fix the parameters $\omega_\text{r,Sq}^0$, $\omega_\text{r,50$\Omega$}$, $\omega_\text{pl,tr}$, $g_\text{tr,Sq}^0$, $g_\text{tr,50$\Omega$}^0 $ for the modelling of the spectroscopy measurements by measuring  the reflectance of the multiplexed SQUID array and redout resonator as a function of SQUID array flux $\Phi_\text{Sq}$, with the DQD far detuned from the SQUID array resonator (Extended Data Fig.~\ref{fig:SampleAndCircuit2}). The blue dots in Extended Data Fig.~\ref{fig:SampleAndCircuit2}(b) are experimentally extracted resonance frequencies from  Extended Data Fig.~\ref{fig:SampleAndCircuit2}(a) and the red dots are the eigenvalues of the calculated system Hamiltonian \eqref{hamilt}.
From the fit we extract the parameters: $\omega_\text{pl,tr}/2\pi = 6.550$ GHz, $\omega_\text{r,Sq}^0/2\pi  = 7.867$ GHz, $\omega_\text{r,50$\Omega$}/2\pi  = 6.490$ GHz, $E_\text{c}/h = 243$ MHz, $\beta = 0.1$, $g_\text{tr,Sq}^0/2\pi  = 230$ MHz, $g_\text{tr,50$\Omega$}^0/2\pi  = 120$ MHz.
We parameterize the total magnetic flux of the SQUID array as: $\Phi_\text{sq}^\prime=2 \pi\gamma \Phi_\text{Sq} + 2 \pi \Phi_\text{c}$, with $\gamma \sim 0.43$ and $\Phi_\text{c} \sim 0.0072 \times \Phi_0$.
The transmon magnetic flux is
$\Phi_\text{tr}^\prime= 2\pi \alpha \Phi_\text{Sq} + 2\pi \Phi_\text{tr}$, with the ratio $\alpha \sim 4.41$ between the areas of the SQUID loops of the transmon and the SQUID array resonator. To fit the data points in Extended Data Figure~\ref{fig:SampleAndCircuit2}, we fixed $\Phi_\text{tr} \sim 0.162 \times \Phi_0$.

We note that the transmon and 50 $\Omega$ resonator are well described by the model for all values of applied flux $\Phi_\text{Sq}$.
The dependence of the SQUID array resonant frequency from $\Phi_\text{Sq}$ is well captured for $-0.3 \leq \Phi_\text{Sq}/\Phi_0 \leq 0.3$. However, for $|\Phi_\text{Sq}| > 0.3\times \Phi_0$ the lumped element description fails to describe the the SQUID array cavity properties accurately. One reason might be due to non-linearity of the Josephson junctions forming the SQUID array resonator, or to inhomogeneity in the magnetic flux threading the SQUIDs of the array. The parameters of the 50 $\Omega$ resonator and transmon are kept fixed for all the other fits. We adjust only parameters related to the SQUID array and DQD, according to the expressions in Eqs.~\eqref{eq8}, \eqref{eq10}, \eqref{eq15}-\eqref{eq17}.

%%%%%%%%%%%%%%%%%%%%%%%%%%%%%%%%%%%%%%%%%%%%%%%%%%

\subsection*{Coherent coupling of DQD to SQUID array and transmon to SQUID array}

\begin{figure}[!h]
\begin{center}
\includegraphics[width=0.50\textwidth]{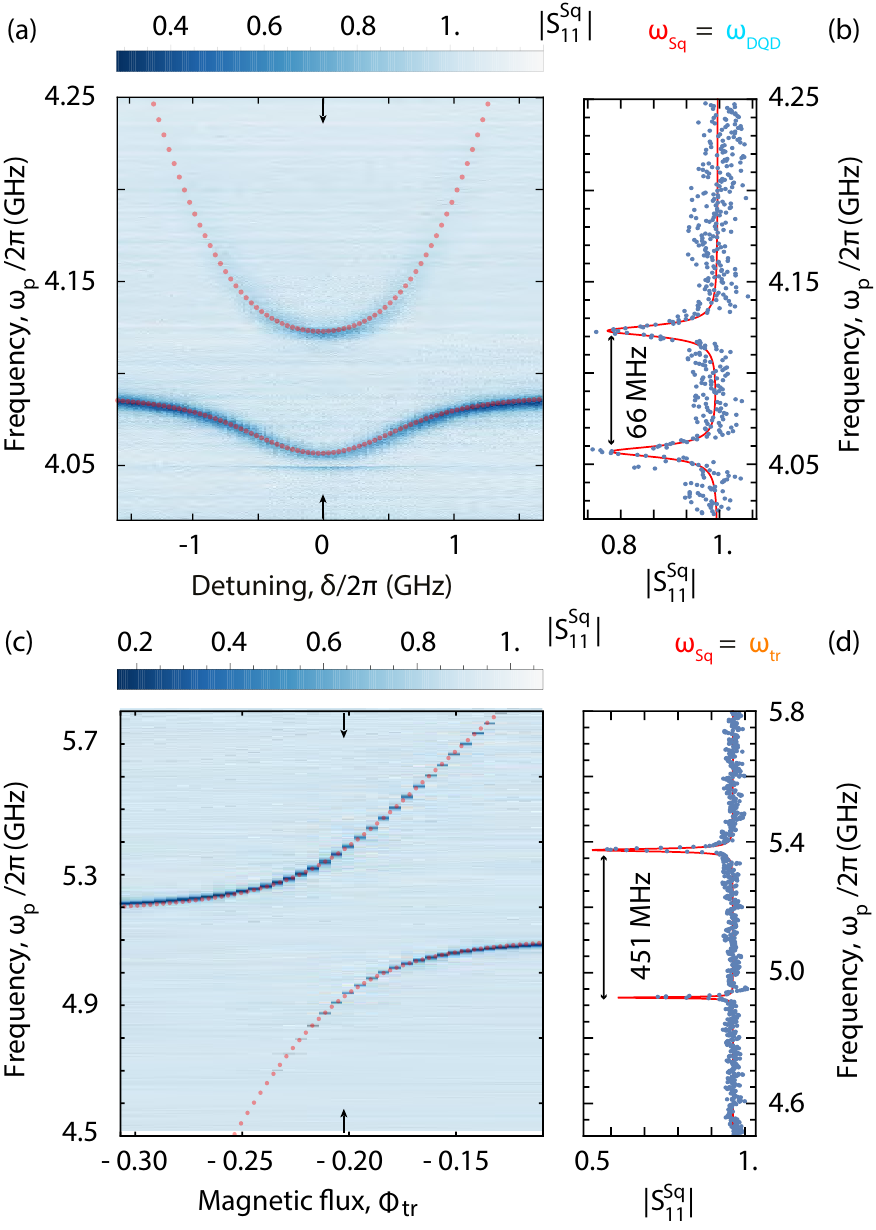}
\end{center}
\caption{
Spectroscopy of the DQD-SQUID array and transmon-SQUID array vacuum Rabi mode splittings.
(a) Reflection response $|S_{11}^{\rm{Sq}}|$ versus detuning $\delta$ of the DQD qubit hybridized with the resonator, resulting in a vacuum Rabi splitting at $\delta=0$. The red points are the results of the systems spectrum simulation, according to the Eq.~\eqref{hamilt} with parameters
$2t_{\rm{c}}/h = 4.090$ GHz, $\omega_\text{r,Sq}/2\pi = 4.089$ GHz, $g_\text{DQD,Sq}/2\pi  = 33$ MHz and far detunned transmon at $\omega_{\rm{tr}}/2\pi \sim 1.720$ GHz.
(b) $|S_{11}^{\rm{Sq}}|$ at $\delta=0$ showing the vacuum Rabi splitting with $2g_{\rm{DQD}}/2\pi \sim 66\,\rm{MHz}$ and an effective linewidth of $\delta \omega_{\text{DQD}}/2\pi \sim7.1\,\rm{MHz}$. The solid line is a fit to a sum of two Lorentzians.
(c) Reflection response $|S_{11}^{\rm{Sq}}|$ versus the applied magnetic flux ($\Phi_\text{tr}$) through the SQUID loop of the transmon qubit hybridized with the SQUID array resonator, resulting in an avoided crossing around $\omega_{\rm{tr}}/ 2 \pi \sim \omega_{\rm{r,Sq}}/ 2 \pi  \sim 5.125\,\rm{GHz}$.
Simulation parameters:
$\omega_\text{r,Sq}/2\pi = 5.181$ GHz, $g_\text{tr,Sq}^0/2\pi  = 271$ MHz [the couplings $g_{\rm{tr},50\Omega}$ and $g_{\rm{tr,Sq}}$ vary with $\Phi_{\rm{tr}}$]. $\omega_{\rm{DQD}}/2\pi \sim 10.8$ GHz ($2t_{\rm{c}}/h \sim 4$ GHz and $\delta/h \sim10$ GHz).
(d) $|S_{11}^{\rm{Sq}}|$ at the flux indicated by the black arrows in panel (c) showing the vacuum Rabi splitting with $2g_{\rm{tr}}/2\pi \sim450\,\rm{MHz}$ and an effective linewidth of $\delta \omega_{\text{tr}}/ 2\pi\sim7.5\,\rm{MHz}$.
}
\label{fig:SampleAndCircuit3b}
\end{figure}

To quantify the coupling strength between the SQUID array resonator and the DQD, we first fix the transmon resonance frequency $\omega_{\rm{tr}}/2\pi< 2 \, \rm{GHz}$, the SQUID array resonance frequency $\omega_{\rm{r,Sq}}/2\pi =4.089  \, \rm{GHz}$ and adjust the tunnel coupling of the DQD to the same frequency [see Extended Data Fig.~\ref{fig:SampleAndCircuit3b}(a)].

With the DQD far detuned, the undercoupled ($\kappa_{\rm{int}}>\kappa_{\rm{ext}}$) SQUID array resonator displays the bare linewidth $\kappa/2\pi=(\kappa_{\rm{ext}}+\kappa_{\rm{int}})/2\pi \sim(4+8)$ MHz.
Varying the DQD detuning $\delta$, we bring the DQD into resonance with the SQUID array resonator ($\omega_\text{r,Sq}= \omega_{\text{DQD}}$) close to the DQD sweetspot, indicated by arrows in Extended Data Fig.~\ref{fig:SampleAndCircuit3b}(a). We observe a clear vacuum Rabi mode splitting in the reflectance spectrum of the resonator [Extended Data Fig.~\ref{fig:SampleAndCircuit3b}(b)].
A fit (red line) of the spectrum to two Lorentzian lines yields a splitting of $2 g_{\rm{DQD,Sq}}/2\pi \sim 66$ MHz, with an effective linewidth $\delta \omega_\text{q} /2 \pi \sim 7.1$ MHz dominated by the loss of the SQUID array resonator.
Keeping the DQD in Coulomb blockade ($\omega_{\rm{DQD}}/2\pi > 10 \, \rm{GHz}$), we tune the transmon frequency $\omega_{\rm{tr}}$ while keeping the SQUID array resonance frequency fixed at $\omega_{\rm{r,Sq}} /2 \pi=5.1813 \, \rm{GHz}$, by using a linear combination of the magnetic flux
generated by a superconducting coil, mounted perpendicular to our sample holder, and a flux line.
We realize the parameter configuration shown in Extended Data Fig.~\ref{fig:SampleAndCircuit3b}(c), with the transmon and the SQUID array in resonance, from which we extract the transmon SQUID array resonator coupling strength $2 g_{\rm{tr,Sq}}/2\pi \sim 451$ MHz.

\subsection*{Simulation of the population transfer in Fig.~\ref{fig:SampleAndCircuit5}(e)}

The Markovian master equation used to fit the time-resolved transmon population oscillations in Fig.~\ref{fig:SampleAndCircuit5}(e), treats both qubits as two-level systems in the dispersive regime with an exchange interaction $2J/2\pi = 21.6\,\text{MHz}$ consistent with the spectroscopically measured energy splitting $2J/2\pi \sim 21\, \rm{MHz}$ [Fig.~\ref{fig:SampleAndCircuit5}(b-c)].
It includes spontaneous emission and dephasing terms for the transmon and a dephasing term for the DQD. We determine the dephasing rates in independent measurements by measuring the DQD charge qubit linewidth $\gamma_2/ 2 \pi=2.6\,\text{MHz}$ and by performing Ramsey measurements of the transmon qubit resulting in $T_2^*=127$ ns. The relaxation rate $T_1=185$ ns of the transmon is used as a fitting parameter in the simulation.

This relaxation rate is close to what expected by considering the Purcell decay ($\Gamma_{\rm{1,tr}} \approx k_{\rm{tot}} g^2_{\rm{tr,Sq}}/\Delta_{\rm{tr,Sq}}^2 \sim 1\, \rm{MHz}$) of the transmon through the SQUID array resonator at the chosen detuning. Additionally, we include a finite time difference $\Delta\tau=23\,\text{ns}$ between the preparation pulse and the flux pulse as a fitting parameter.
The flux pulse is simulated as a square pulse subject to a Gaussian filter with standard deviation $\sigma=3\,\text{ns}$. In the experiment, the finite rise time of the flux pulse results most likely from the finite bandwidth (300 MHz) of the employed arbitrary wave form generator (Tektronix 5014 AWG).

\end{document}